\documentstyle[12pt,epsf]{article}
 
\newcommand{\BEQ}{\begin{equation}}    
\newcommand{\BEA}{\begin{eqnarray}}
\newcommand{\EEQ}{\end{equation}}      
\newcommand{\EEA}{\end{eqnarray}}
\newcommand{\rar}{\rightarrow}                   
\newcommand{\zeile}[1]{\vskip #1 \baselineskip}  

\newcommand{\build}[3]{\mathrel{\mathop{\kern 0pt#1}\limits_{#2}^{#3} }}

\catcode`\@=11
\def\numberbysection{\@addtoreset{equation}{section}
        \def\theequation{\thesection.\arabic{equation}}}

\begin{document}
\baselineskip 0.3in
%
%
%
\begin{titlepage}
\begin{center}
{\Large \bf Persistent currents in mesoscopic rings and conformal 
invariance\footnote{Dedicated to Prof. J. Zittartz on the occasion of his
60$^{\rm th}$ birthday.}}
\vskip 0.5in
Malte Henkel and Dragi Karevski 
 \\[.3in]
{\em Laboratoire de Physique des Mat\'eriaux,\footnote{Unit\'e Mixte de
Recherche CNRS No. 7556.} \\ Universit\'e Henri Poincar\'e
Nancy I, B.P. 239, \\ 
F - 54506 Vand{\oe}uvre-l\`es-Nancy Cedex, France} \\
\end{center}
\zeile{1}
%
\begin{abstract}
The effect of point defects on persistent currents in mesoscopic
rings is studied in a simple tight-binding model. Using an analogy
with the treatment of the critical quantum Ising chain with defects, 
conformal invariance techniques are employed to relate the persistent 
current amplitude to the Hamiltonian spectrum just above the Fermi energy. 
{}From this, the dependence of the current on the magnetic flux is found exactly
for a ring with one or two point defects. 

The effect of an aperiodic modulation of the ring, generated through a binary 
substitution sequence, on the persistent
current is also studied. The flux-dependence of the current is found
to vary remarkably between the Fibonacci and the Thue-Morse sequences. 
\end{abstract}
\zeile{1}
\end{titlepage}

\newpage
The study of the critical behaviour of spin systems close to surfaces
and interfaces has revealed a very rich phenomenology and has provided much
insight into new phenomena. Two-dimensional systems have
been of particular theoretical interest, because of fruitful connections with 
conformal invariance techniques, see \cite{Igl93,Car96,diF97,Hen98} for an
introduction and review. Here, we present a new application
of some conformal invariance methods to a simple problem of persistent 
currents, which we now specify. 

Consider a small ring, made from a semiconducting or a normally conducting
material. If the ring is threaded by a magnetic flux, one observes
experimentally a persistent current, provided the ring has mesoscopic
dimensions, that is, its diameter is in the nanometer range. Measured
examples include Au \cite{E:Cha91} or GaAS-AlGaAs \cite{E:Mai93} rings. 
For larger rings, the current decreases and is no longer perceptible for
rings of macroscopic size. The problem has received much attention.
Non-interacting rings were studied in the classic papers by Kulik \cite{kulik70}
and B\"uttiker {\it et al.} \cite{butti83}. 
Disordered rings were considered in \cite{cheung89}. 
On the other hand, persistent currents for interacting
$1D$ electrons were first investigated in \cite{zvy90,shas90}.
See also \cite{Gog93,Kri96,Poi94,Abr93,Aff92,Vol97,Qin97,Los92} 
and references therein. 
We refer to \cite{zvy95} for a review and to the nice book by 
Imry \cite{Imr97} for a 
detailed presentation of the physical background. 

\begin{figure}
\centerline{\epsfxsize=2.5in\epsfbox
{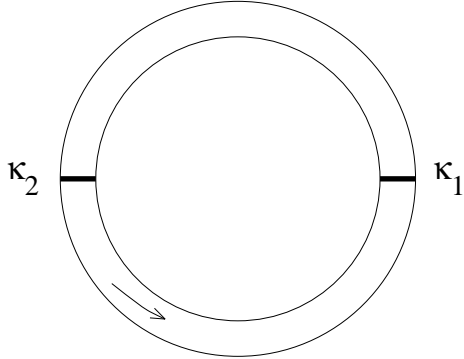}}
\caption[Aharonov-Bohm geometry]{Aharonov-Bohm geometry. The places of the 
defects and their couplings $\kappa_{1,2}$ are indicated. The magnetic field
is perpendicular to the plane. 
\label{fig:Dauerstrom}}
\end{figure}

Here, we shall limit ourselves to a simple toy model which describes 
non-interacting (spin-less) electrons moving around a small, thin ring
penetrated by an Aharonov-Bohm flux $\Phi$, see Figure~\ref{fig:Dauerstrom}.
The flux $\Phi=\oint A dx = A L$, where $A$ is the vector
potential and $L$ is the length of the ring. 
Assume that the ring is discretized
into $N$ cells of size $a$. We use a simple 
{tight-binding model} which describes
the hopping of electrons from one cell to the next
\BEQ \label{eq:HuepfHamFluss}
H = -t_{\rm hop} \sum_{n=1}^{N} \left[ \exp \left( 2\pi i \frac{\Phi}{\Phi_0}
\frac{a}{L} \right) \psi^{\dag}(n+1) \psi(n) + \mbox{\rm h.c.} \right]
\EEQ
where $t_{\rm hop}$ is the hopping rate and 
$\Phi_0=hc/e$ is the flux quantum. Hamiltonians of this kind have been 
proposed earlier, see \cite{shas90,Los92}, to study the effect of 
an Aharonov-Bohm flux on
a Luttinger liquid. Our model (\ref{eq:HuepfHamFluss}) is, with respect to
\cite{shas90,Los92}, simplified even further in neglecting
the charge-density interactions and only considering the zero-temperature
case. On the other hand, we allow for point-like defects to occur in the 
ring, which
phenomenologically lead to a local modification of the hopping rate, as
already studied in \cite{Gog93}. 

We shall limit ourselves throughout to the case
of {half-filling}, when exactly $M=N/2$ particles are present and assume 
that $N$ is even. The flux $\Phi$ can be related to the
phase 
\BEQ
\phi=2\pi \Phi/\Phi_0
\EEQ
of the wave function through the gauge transformation
\BEQ \label{Gl:EichTrans}
\psi(n) \rar \exp\left( i\phi \frac{a}{L} n \right) \psi(n)
\EEQ
leading to the conventional tight-binding Hamiltonian
\BEQ \label{eq:EnggebundenHam}
H = -t_{\rm hop} \sum_{n=1}^{N} \left[  
\psi^{\dag}(n+1) \psi(n) +\psi^{\dag}(n)\psi(n+1) \right]
\EEQ
with the boundary condition $\psi(N+1)=e^{-i\phi}\psi(1)$ and where $N$ is 
the number of sites. This is nothing
but the classical Byers-Yang-Bloch result that the system is periodic in the
flux $\Phi$ with period $\Phi_0$, see \cite{shas90,Los92,zvy95,Imr97} 
and references therein. In particular, the amplitude $j$ of the persistent 
current is obtained from the ground 
state energy $E_0(\Phi)$ of $H$ as 
\BEQ \label{Gl:Stromdichte}
j = - \frac{\partial E_0}{\partial \Phi}
\EEQ
It is well known that for a perfect ring, 
$j=(t_{\rm hop}/\pi)(\phi/N)$ inside the period interval
$-\pi<\phi\leq \pi$ and continued periodically outside it 
\cite{zvy95,Imr97}.\footnote{To facilitate comparison between
the models considered here, signs and phases are arranged such as to reproduce 
this result for the defectless case throughout.} 
Note the scaling of $j$ with the ring length $L=Na$ \cite{zvy95,Imr97}, 
so that the current vanishes in the bulk limit $L\rar\infty$, as expected.   

We now consider the situation where in an otherwise perfect ring there is a
single link where the hopping rate 
is $t_{\rm hop}\rho$,\footnote{The impurities 
studied here are only forward scattering.} 
rather than $t_{\rm hop}$, see Figure~\ref{fig:Dauerstrom}. By rotational
symmetry, we thus have a defect boundary condition $\psi(N+1)=\kappa\psi(1)$ 
with a {\em complex} defect strength $\kappa = \rho e^{-i\phi}$. Formally,
the diagonalization of the 
free fermion $XX$ Hamiltonian (\ref{eq:EnggebundenHam}) proceeds along exactly 
the
same lines as had been developed earlier for the treatment of the quantum
Ising chain in the presence of point defects (or, equivalently, the classical
$2D$ Ising model with line defects) \cite{Igl93,Hen87,Hen89a}. 
Through a Jordan-Wigner
transformation and a subsequent canonical transformation, $H$ can be written
in terms of free fermionic harmonic oscillators $\eta_k$ \cite{lieb}
\BEQ
H = \sum_{k} \Lambda_k \left(\eta_k^{\dag}\eta_k -\frac{1}{2}\right)
\EEQ
where $\Lambda_k=-t_{\rm hop}\cos k$ and the values of $k$ are 
determined by solving
\BEQ \label{Gl:KBedingung}
\rho^2 \sin( k(N-1) ) +2\rho\cos\phi \sin k -\sin(k(N+1)) =0
\EEQ
The physical ground state 
is obtained by filling up the lowest $M=N/2$ one-fermion states. Formally,
the ground state energy $-\frac{1}{2} \sum_k \Lambda_k$ can now be
calculated and the current $j$ be obtained. Practically, this is not 
straightforward, since there is no closed solution to (\ref{Gl:KBedingung})
for $|\rho|\neq 1$. 

However, to a collective effect as a persistent current
only the lowest excited levels should make a significant contribution. Plotting
$\Lambda_k$ as a function of $k$, one sees that for half-filling, the lowest
modes occur at values of $k$ around $\frac{\pi}{2}$ and $\frac{3\pi}{2}$. 
Linearizing $\Lambda_k$ close to the Fermi energy, we can therefore replace the
exact $H$ by an effective low-energy Hamiltonian $H_{\rm eff}$ of the form 
\BEQ \label{Gl:effektHam}
H_{\rm eff} = \frac{2\pi t_{\rm hop}}{N} \sum_{r=0}^{\infty} \left[ \left( r 
+\frac{1}{2} - 
{\cal D} \right) n_{r}^{(-)} + \left( r +\frac{1}{2} + 
{\cal D} \right) n_{r}^{(+)} \right] - \frac{t_{\rm hop}\pi c_{\rm eff}}{6 N} 
+{\cal O}(N^{-2})
\EEQ
where $n_{r}^{(\pm)}$ are fermionic number operators representing the two
lowest modes and where the one-fermion 
energy shift is ${\cal D}={\cal D}(\rho,\phi)$, 
such that ${\cal D}(1,0)=0$. The
same form had been found before for the Ising quantum chain with 
defects \cite{Hen87,Hen89a}. 

In (\ref{Gl:effektHam}) we have dropped a constant 
contribution of the form $F N + S$. However, these terms have no bearing on the
persistent current. To see this we use the analogy of the $XX$ chain 
(\ref{eq:EnggebundenHam})
with the Ising quantum chain. In turn, the Ising quantum chain can be viewed as 
the Hamiltonian limit of a two-dimensional classical Ising system, making
contact with standard thermodynamics. 
From this, we recognize $F$ as the bulk free energy density per site 
and $S$ corresponds to a surface free energy term.
This interpretation is carried over to the $XX$ chain. Since the bulk free 
energy density $F$ is independent of any
boundary condition, it is thus also independent on the {\em complex} 
defect strength $\kappa$. On the
other hand, in the same framework the surface free energy $S$ may be
sensitive to boundary effects. Because of the gauge transformation 
(\ref{Gl:EichTrans}), the flux $\phi$ can also be viewed as a bulk interaction,
and it follows that $S$ will merely depend on the {\em real} 
defect strength $\rho$, but
not on $\phi$. In conclusion, since $F,S$ are both independent of $\phi$, 
eq.~(\ref{Gl:effektHam}) contains all the information required for the 
determination of the persistent current through (\ref{Gl:Stromdichte}). 
 
Because the dispersion relation close to the Fermi energies is {\em linear},
the effective Hamiltonian represents a conformally 
invariant system \cite{Car96,diF97,Hen98}. 
Conformal invariance 
allows to relate the (half-filled) physical ground state energy
\BEQ
E_0 = \sum_{-{\pi\over 2} < k \leq {\pi\over 2}} \Lambda_k = 2 E_{0,{\rm eff}}
\EEQ
to the constant $c_{\rm eff}$ and furthermore to the
one-fermion energy shift $\cal D$.
Here, $E_{0,{\rm eff}}$ is the ground state energy of $H_{\rm eff}$ and the
factor 2 comes from the symmetry of $\Lambda_k$ at half-filling. 

One of the main results of $2D$ conformal invariance 
(e.g. \cite{Car96,diF97,Hen98}), 
applied to critical classical spin sytems, 
states that it acts as a dynamical symmetry for the associated quantum chain 
of $N$ sites, with Hamiltonian
\BEQ \label{Gl:DynaSymm}
H = \frac{2\pi}{N} \left( L_0 + \bar{L}_0 \right) - \frac{\pi c}{6N}
\EEQ
where $L_0,\bar{L}_0$ are generators of the Virasoro algebra and $c$ is the
central charge. Here, the quantum Hamiltonian $H$ is obtained from the
transfer matrix ${\cal T}=\exp(-\tau H)$ of the classical $2D$ spin model, see
\cite{Igl93,diF97,Hen98}. 
As it stands, (\ref{Gl:DynaSymm}) applies to systems without
any defects at all and in particular to quantum chains with periodic boundary 
conditions. One may easily recover the spectrum-generating
conformal algebra for the excitations described by $H_{\rm eff}$ 
above the Fermi threshold, in complete
analogy with the familiar construction for critical Ising spin systems, see
\cite{diF97,Hen98}. 

Now, introducing non-periodic boundary conditions (which
may also be viewed as a point defect in an otherwise periodic quantum chain), 
it can be shown that
the transformations generated by the diagonal subalgebra $K_n = L_n +\bar{L}_n$
are preserved while the other generators $L_n - \bar{L}_n$ no longer correspond
to a symmetry of the system. Such a point defect in the quantum chain
Hamiltonian corresponds, via a conformal transformation, to a semi-infinite
line defect in the associated $2D$ spin model, see \cite{Igl93,Hen98} and
references therein. Specifically, for the Ising quantum chain 
it turns out that the
perfect generators $L_n$ are modified into Virasoro generators $L_n({\cal D})$ 
depending explicitly on the one-fermion shifts such that \cite{Hen87}
\BEQ \label{eq:KHam}
K_n = K_n({\cal D}) = L_n({\cal D})+\bar{L}_n({\cal D})+
\frac{1}{2}{\cal D}^2 \delta_{n,0}
\EEQ
satisfy a Virasoro algebra with central charge $c=1$ and
\BEQ
H_{\rm eff} = \frac{2\pi}{N} \left( K_0({\cal D}) - \frac{1}{24} \right) 
\EEQ
which generalizes (\ref{Gl:DynaSymm}). Although invariance under the full
conformal algebra generated by the set $\{L_n,\bar{L}_n\}$ is broken, the 
diagonal conformal subalgebra generated by the $K_n$ still acts as a dynamical
symmetry. Thus, $c_{\rm eff} = \frac{1}{2}-6{\cal D}^2$ \cite{Hen87} and 
the only change in the ground state
energy of $H$ which can give rise to a persistent current is
\BEQ
\delta E_0 = 2\delta E_{0,{\rm eff}} =(2\pi/N) t_{\rm hop} {\cal D}^2
\EEQ 
This is the desired result, which relates the current $j$ to the shift $\cal D$
of the lowest excitations. 

\begin{figure}
\centerline{\epsfxsize=5.75in\epsfbox
{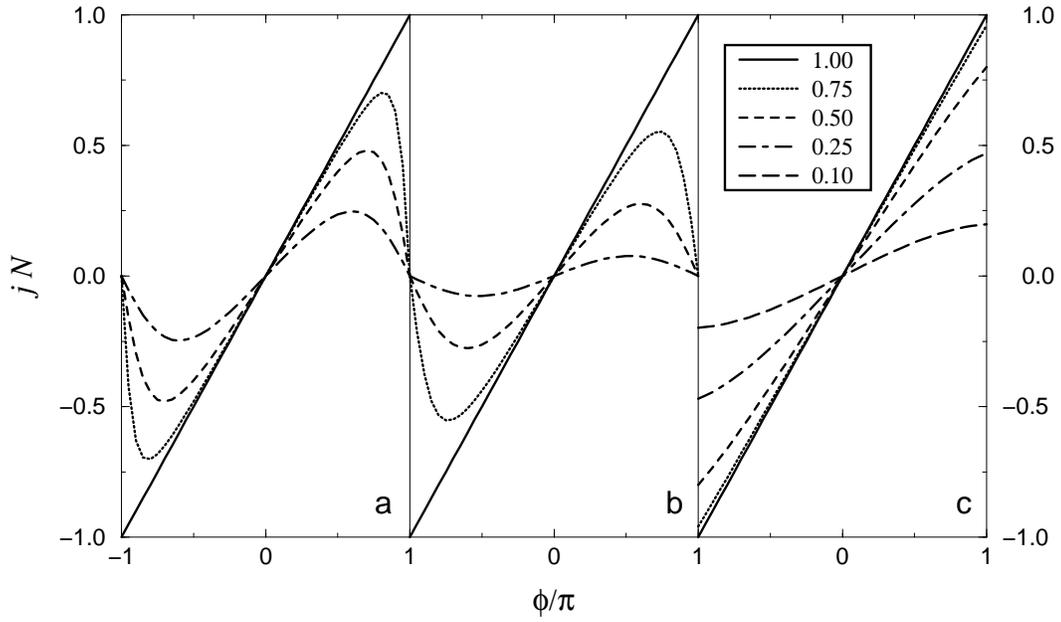}}
\caption[Persistent current amplitude in the tight-binding model]{Persistent
current amplitude $j(\phi) N$ in the tight-binding model for 
(a) a single defect with defect
strength $\kappa=\rho e^{-i\phi}$ for $M=N/2$ odd. For two opposite 
defects with strengths $\kappa_1=\rho e^{-i\phi}$ and $\kappa_2=\rho$ the 
persistent current is shown for (b) $N/2$ even and (c) $N/2$ odd. The 
curves correspond to different values of $\rho$ as indicated 
and $t_{\rm hop}=1$. 
\label{fig3:Dauerstrom}}
\end{figure}

To calculate $\cal D$ to leading order in $1/N$, we first 
shift $k$ to $k+\pi/2$ in order
to have $k=0$ at the Fermi surface and then let $k=\alpha/N$ in
(\ref{Gl:KBedingung}). This gives 
$\cos\alpha = (-1)^M[2\rho/(1+\rho^2)]\cos\phi$. From (\ref{Gl:effektHam}),
we also have $k\simeq (2\pi/N)(\ell+1/2\pm{\cal D})$, 
with $\ell=0,1,2,\ldots$.
Using these two formulas for $k$, we find ${\cal D}={\cal D}(\rho,\phi)$. 
{}From (\ref{Gl:Stromdichte}) the current
$j=-\partial(\delta E_0)/\partial \phi$ is obtained as a function of the phase 
$\phi$. 
Explicitly,
\BEQ \label{Gl:1Defekt}
j(\phi) = t_{\rm hop} \frac{u \arccos( u\cos \phi) \sin\phi}
{\pi\sqrt{1-u^2\cos^2\phi}} \frac{1}{N} \;\; , \;\; 
u = \frac{2\rho}{1+\rho^2}
\EEQ
and we find again the scaling $j\sim N^{-1}$. Note that this expression is
invariant under the transformation $\rho\rar 1/\rho$ which reflects that
the defect, even for $\rho>1$, disturbs the coherence of the many-particle
system. 
As it stands, (\ref{Gl:1Defekt}) is correct for $M$ odd. For $M$ even, 
merely replace $\phi\rar\phi-\pi$. The size-independent current amplitude
$j(\phi) N$ is shown in Figure~\ref{fig3:Dauerstrom}a
for an entire period of the phase $\phi$ and for several values of the 
defect coupling $\rho$. While the periodicity with $\phi$ is kept
unchanged with respect to the perfect case, the amplitude decreases rapidly
with increasing values of $|\rho-1|$. Of course, this result can be checked
through direct numerical diagonalization of the tight-binding
Hamiltonian (\ref{eq:EnggebundenHam}) and may also be reproduced from a simple
continuum description of the model \cite{Gog93}. 

For two defects which are placed opposite to each 
other (Figure~\ref{fig:Dauerstrom}), we find in an analogous fashion
\BEA
j_{\pm}(\phi) &=& t_{\rm hop}\frac{ a \arccos \left( a \cos\phi \pm b\right)
\sin \phi}{\pi\sqrt{ 1 - (a\cos\phi \pm b)^2 }} \frac{1}{N} \nonumber \\
a &=& \frac{4\rho_1\rho_2}{(1+\rho_1^2)(1+\rho_2^2)} \;\; , \;\;
b = \frac{(1-\rho_1^2)(1-\rho_2^2)}{(1+\rho_1^2)(1+\rho_2^2)}
\EEA
where $+$ corresponds to $N/2$ even, $-$ corresponds to $N/2$ odd and
$\rho_{1,2}$ are the couplings at the defects. For $\rho_1=1$ or $\rho_2=1$,
one simply recovers (\ref{Gl:1Defekt}). In Figure~\ref{fig3:Dauerstrom}bc,
we show the current amplitude for the number of particles 
$M=N/2$ even and odd, respectively and for $\rho_1=\rho_2=\rho$.  
In contrast to the single-defect case, there is a 
clear {parity effect}. 
For an {\em even} number of particles, the amplitude is broadly similar to the 
single-defect case (up to a renormalization of $\rho$). In particular, the
current $j(\phi)$ is a continuous function of $\phi$ over the entire
period. For an {\em odd} number
of particles, the flux-dependence of the current 
is almost {\em linear} for a wide
range of values of $\rho$ and there is a jump in $j$ at the ends of the period
interval. Also, the presence of the defects merely changes the
prefactor. 

We now consider the case of two defects placed anywhere on the ring. The first
one may always considered to be placed at the origin, while the second one
will sit at the site $R=\lambda N$. The continuum dispersion relation
remains unchanged, but (\ref{Gl:KBedingung}) has to be replaced by 
\BEQ \label{Gl:K2Bedingung}
\left( 1 - \rho_1^2\right)\left( 1 - \rho_2^2\right) 
\cos\left[ \alpha (1-2\lambda)\right]
\pm \left( 1 + \rho_1^2\right)\left( 1 + \rho_2^2\right) \cos \alpha 
\pm 4\rho_1\rho_2 (-1)^M \cos \phi = 0 
\EEQ
where $N=2M$ is the number of sites, 
$k=\alpha/N$ (after the shift $k\rar k+\pi/2$)
and $+(-)$ corresponds to $R$ even (odd). 
Since $\lambda=R/N$ is rational, we may write $\lambda=p/q$ with $p,q$
mutually prime. Eq.~(\ref{Gl:K2Bedingung}) then becomes a polynomial equation
of degree $\gamma(q)$ in the variable 
$\cos(2\alpha/q)=\cos(2\pi {\cal D}^{(i)})$, where $\gamma(q)=q/2$ if $q$ is 
even and $\gamma(q)=q$ if $q$ is odd. 
The current is then written in terms of the $\gamma(q)$ distinct one-fermion
shifts as
\BEQ
j = -\frac{\partial E_0}{\partial \phi} = -
\frac{4\pi \gamma(q)}{N} \sum_{i=1}^{\gamma(q)}
{\cal D}^{(i)}\frac{\partial {\cal D}^{(i)}}{\partial \phi}
\EEQ
It is straightforward to extend the treatment to an arbitrary number of
point defects, following the lines of \cite{Hen89a} for the Ising quantum 
chain. 

These results were derived by going back to the explicit diagonalization
of the underlying tight-binding model and using the analogy with the
quantum Ising chain for defect boundary conditions. In the Ising quantum
chain, a defect boundary condition corresponds to a marginal perturbation 
\cite{Igl93} which leads to a shift $\cal D$ in the one-fermion energies which 
depends continuously on the defect strengths and consequently 
to a continuously varying  
effective central charge $c_{\rm eff}$.   
Extending existing field-theoretical treatments of defect lines in $2D$
models \cite{Del94,Kon97,Osh97} to complex defect strengths might provide
further insight into the problem at hand.\footnote{For example, the generators
(\ref{eq:KHam}) can also be obtained using the automorphisms of the underlying
Kac-Moody algebra \cite{Baa89}. This more abstract construction is independent
of a free-fermion realization of the model.} 

Having seen that at least in the tight-binding model with point defects, the
persistent currents are related to the shift in the low-lying excitations above
the Fermi energy, we now briefly discuss persistent currents in aperiodically 
modulated rings threaded by a magnetic flux. The Hamiltonian is
\BEQ \label{eq:EnggebundenAperHam}
H = - \sum_{n=1}^{N} t_{n} \left[  
\psi^{\dag}(n+1) \psi(n) +\psi^{\dag}(n)\psi(n+1) \right]
\EEQ
at half-filling, together with the flux boundary condition used before and 
where the aperiodic modulation in the hopping matrix element $t_n$ is 
generated through a binary substitution sequence \cite{Que87,Dum90}. 
We shall consider here 
\BEA
\mbox{\rm Fibonacci : } & ~ & A \rar AB \;\; , \;\; B \rar A \nonumber \\
\mbox{\rm Thue-Morse : } & ~ & A \rar AB \;\; , \;\; B \rar BA 
\EEA
and the sequences are obtained by starting from the letter $A$ and iterating
the above procedure many times. For example, using the Thue-Morse substitution
rules, the first three iterations give the following sequences
\BEQ \begin{array}{rc}
 & A   \\
i=1: & A B  \\
i=2: & A B B A  \\
i=3: & A B B A B A A B \end{array}
\EEQ
A modulation in (\ref{eq:EnggebundenAperHam})
is obtained by letting $t_n = t_{\rm hop}$ if the site $n$ is the letter $A$ 
and $t_n = t_{\rm hop}\rho$ otherwise. For such aperiodic modulation the 
cumulated deviation
from the average hopping rate $\bar{t}$, 
$\Delta(L):=\sum_{k=1}^{L}(t_{k}-\bar{t})$
scales with the chain length $L$ as $\Delta(L)\sim L^{\omega}$, where $\omega$ 
is the so-called wandering exponent, 
caracteristic of the sequence \cite{Que87,Dum90}. If $\omega<0$, 
as it is the case for the two sequences under consideration in this paper, 
$\Delta(L)$ remains bounded for $L\rightarrow\infty$.

For critical spin systems, Luck \cite{Luc93}
has formulated a criterion which for a given spin system and a given
sequence predicts whether the aperiodic modulation is relevant, marginal
or irrelevant. According to the Luck criterion, the modulations generated by 
both the Fibonacci and the
Thue-Morse sequence are irrelevant for the Ising quantum chain and the 
low-lying excitations of $H$ are in the $N\rar\infty$ limit not affected by the
modulation. Because of the close connection of the tight-binding model
with the critical point of the Ising quantum chain, 
we expect to retain the scaling $j\sim N^{-1}$ of the
persistent current, but we are curious about the flux-dependence of $j(\phi)$. 

\begin{figure}
\epsfxsize=3.75in
\centerline{\epsfbox
{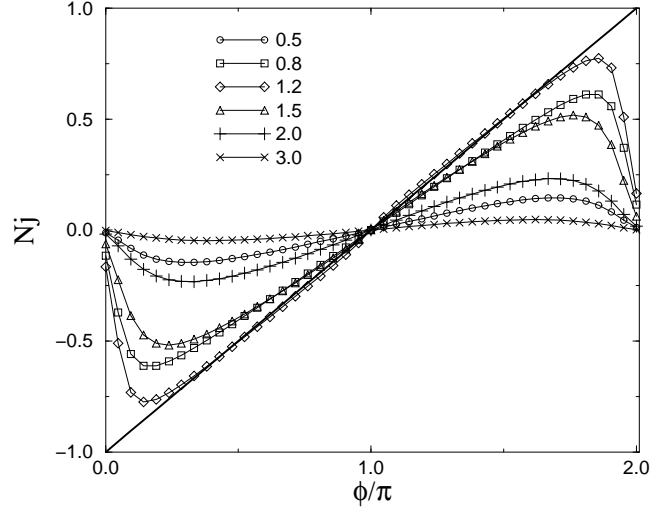}}
\caption[FiboN]{Current amplitude $N j$ for a ring modulated with the
Fibonacci sequence, $N=144$, $t_{\rm hop}=1$ 
and different values of $\rho$. \label{fig:fibor}}
\end{figure}

\begin{figure}
\epsfxsize=3.75in
\centerline{\epsfbox
{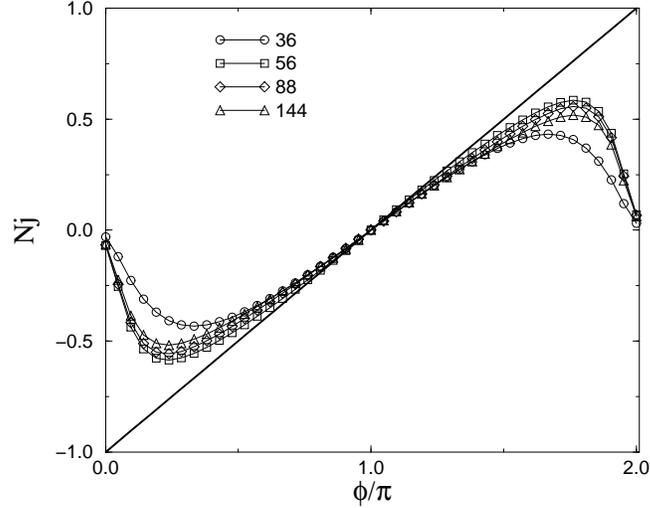}}
\caption[Fibor]{Current amplitude $Nj$ for a ring modulated with the
Fibonacci sequence, $\rho=1.5$ and varying number of sites $N$. 
\label{fig:fibon}}
\end{figure}

\begin{figure}
\epsfxsize=3.75in
\centerline{\epsfbox
{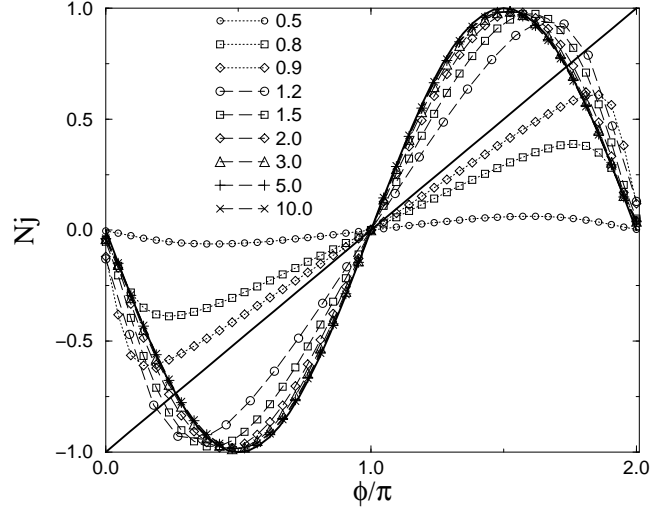}}
\caption[ThueMorseN]{Current amplitude $Nj$ for a ring modulated with the
Thue-Morse sequence, $N=64$, $t_{\rm hop}=1$ and several values of $\rho$. 
\label{fig:thuemorse}}
\end{figure}

In Figure~\ref{fig:fibor}, 
we show the current amplitude $jN$ as obtained
for several values of $\rho$ for an aperiodic ring modulated with the
Fibonacci sequence. We also include the straight line behaviour for the
unmodulated ring for comparison. We observe that the modulation reduces
the current density with respect to the unmodulated case and that
the dependence of $j(\phi)$ on the flux is similar to the one found for the
single defect in Figure~\ref{fig3:Dauerstrom}a. In Figure~\ref{fig:fibon},   
we illustrate for fixed $\rho$
the convergence of the finite-size data with $N$. We see that quite large
lattices are needed to achieve convergence of the current amplitude. 
Finally, in 
Figure~\ref{fig:thuemorse}
the current amplitudes for the Thue-Morse sequence are displayed. In this
case, the convergence with $N$ is much faster than for the Fibonacci sequence 
and already for $N=64$, there is no
perceptible change of the amplitude with $N$. 

Remarkably, the current depends on the flux in very different ways for the
two sequences, although we observed indeed the expected scaling $j\sim N^{-1}$. 
For the Fibonacci sequence, the 
amplitude $N j(\phi)$ with $\rho\neq 1$ is always smaller 
than the amplitude for 
the  non-modulated ring $\rho=1$, independently of whether
$\rho$ is larger or smaller than unity. That is not so for the Thue-Morse
sequence. Rather, the amplitude $N j$ is weakened for $\rho<1$, 
but for $\rho>1$,
the amplitude becomes considerably larger than in the unmodulated case. 
In fact, for $\rho\rar\infty$, 
we recognize a limit form $N j(\phi) = -\sin \phi$. In spite of being
irrelevant to the critical behaviour of Ising quantum chains, the details
of the aperiodic modulation are important for the flux-dependence of the
persistent current. These finer properties are not reflected in the low-lying
excitation spectrum of the Hamiltonian.

In conclusion, the role of defects on the persistent current in mesoscopic
rings was studied through a simple tight-binding model. From the analogy
with the local critical behaviour of conformally invariant spin systems 
with defect lines, when extended to complex defect strengths, explicit 
formulae for the persistent current were obtained. The flux-dependence
of the current shows an unexpected parity effect. 
In addition, the persistent current was found for two types of aperiodically 
modulated rings. The flux-dependence of the current $j(\phi)$ cannot be 
obtained from a consideration of the low-lying excitation spectrum alone, in 
contrast to the case of point defects. 
It would be interesting to see to what extent these observations can be
generalized beyond the simple tight-binding model.

\zeile{1}
\noindent {\bf Acknowledgement:} 
It is a pleasure to thank H. Schoeller for very useful
discussions. 

 

{\small  
\begin{thebibliography}{99}
\bibitem{Igl93} F. Igl\'oi, I. Peschel and L. Turban, Adv. Phys. {\bf 42}, 
683 (1993)
\bibitem{Car96} J.L. Cardy, {\it Scaling and Renormalization in Statistical
Physics}, Cambridge University Press (Cambridge 1996)
\bibitem{diF97} P. di Francesco, P. Mathieu and D. S\'en\'echal, {\it
Conformal Field Theory}, Springer (Heidelberg 1997)
\bibitem{Hen98} M. Henkel, {\it Conformal Invariance and Critical Phenomena},
Springer (Heidelberg 1998), Chap 15
\bibitem{E:Cha91} V. Chadrasekhar, R.A. Webb, M.J. Brady, M.B. Ketchen,
W.J. Gallagher and A. Kleinsasser, Phys. Rev. Lett. {\bf 67}, 3578 (1991)
\bibitem{E:Mai93} D. Mailly, C. Chapelier and A. Benoit, Phys. Rev.
Lett. {\bf 70}, 2020 (1993)
\bibitem{kulik70} I.O. Kulik, JETP Lett. {\bf 11}, 275 (1970)
\bibitem{butti83} M. B\"uttiker, Y. Imry and R. Landauer, Phys. Lett. {\bf 96A}, 
365 (1983)
\bibitem{cheung89} H.F. Cheung and E.K. Riedel, Phys. Rev. {\bf B40}, 
9498 (1989)
\bibitem{zvy90} A.A. Zvyagin, Sov. Phys. Solid State, {\bf 32}, 905 (1990)
\bibitem{shas90} B.S. Shastry and B. Sutherland, Phys. Rev. Lett. {\bf 65}, 243 
(1990)
\bibitem{Gog93} A.O. Gogolin, Phys. Rev. Lett. {\bf 71}, 2995 (1993); 
A.O. Gogolin and N.V. Prokofev, Phys. Rev. {\bf B50}, 4921 (1994)
\bibitem{Kri96} I.V. Krive, P. Sandstr\"om, R.I. Shekhter and M. Jonson, 
Phys. Rev. {\bf B54}, 10342 (1996); 
P. Sandst\"om and I.V. Krive, Ann. of Phys. {\bf 257}, 18 (1997)
\bibitem{Poi94} G. Bouzerar, D. Poilblanc and G. Montamboux, 
Phys. Rev. {\bf B49}, 8258 (1994); {\bf B52}, 10772 (1995); 
F. Mila and D. Poilblanc, Phys. Rev. Lett. {\bf 76}, 287 (1996)
\bibitem{Abr93} M. Abraham and R. Berkovits, Phys. Rev. Lett. {\bf 70}, 1509
(1993)
\bibitem{Aff92} E.S. S{\o}rensen, S. Eggert and I. Affleck, J. Phys. {\bf A26}, 
6757 (1993);
S.J. Qin, M.Fabrizio, L. Yu, M. Oshikawa and I. Affleck, Phys. Rev. {\bf B56},
9766 (1997)
\bibitem{Vol97} A. V\"olker and P. Kopietz, Z. Phys. {\bf B102}, 545 (1997);
Mod. Phys. Lett. {\bf B10}, 1397 (1997)
\bibitem{Qin97} S.J. Qin, M. Fabrizio and L. Yu, Phys. Rev. {\bf B54}, R9643
(1996) 
\bibitem{Los92} D. Loss, Phys. Rev. Lett. {\bf 69}, 343 (1992)
\bibitem{zvy95} A.A. Zvyagin and I.V. Krive, Low Temp. Phys. {\bf 21}, 533 
(1995)
\bibitem{Imr97} Y. Imry, {\it Introduction to Mesoscopic Physics}, 
Oxford University Press (Oxford 1997)  
\bibitem{Hen87} M. Henkel and A. Patk\'os, Nucl. Phys. {\bf B285}, 29 (1987)
\bibitem{Hen89a} M. Henkel, A. Patk\'os and M. Schlottmann, Nucl. Phys. 
{\bf B314}, 609 (1989)
\bibitem{lieb} One can find the explicit derivation in the classical paper
E. Lieb, T. Schultz and D. Mattis, Ann. of Phys. {\bf 16}, 407 (1961)
\bibitem{Del94} G. Delfino, G. Mussardo and P. Simonetti, Nucl. Phys. 
{\bf B432}, 518 (1994)
\bibitem{Kon97} R. Konik and A. LeClair, preprint {\tt hep-th/9703085}
\bibitem{Osh97} M. Oshikawa and I. Affleck, Nucl. Phys. {\bf B495}, 533 (1997)
\bibitem{Baa89} M. Baake, P. Chaselon and M. Schlottmann, Nucl. Phys. 
{\bf B314}, 625 (1989)
\bibitem{Que87} M. Qu\'effelec, in A. Dold and B. Eckmann (Eds) {\it
Substitution Dynamical Systems - Spectral Analysis},  Lecture Notes in 
Math., vol. 1294, Springer (Heidelberg 1987), p. 97
\bibitem{Dum90} J.M. Dumont, in J.M.Luck, P. Moussa and M. Waldschidt (eds) 
{\it Number Theory and Physics}, Springer (Heidelberg 1990), p. 185
\bibitem{Luc93} J.M. Luck, J. Stat. Phys. {\bf 72}, 417 (1993); Europhys. Lett.
{\bf 24}, 359 (1993) 

\end{thebibliography} 
}
\end{document}